\documentclass[12pt]{article}

\DeclareSymbolFont{bletters}{OML}{cmm}{bx}{it}
\DeclareMathSymbol{\bsigma}{\mathord}{bletters}{"1B}
\DeclareMathSymbol{\bphi}{\mathord}{bletters}{"1E}
\DeclareMathSymbol{\bbe}{\mathord}{bletters}{"0C}
\DeclareMathSymbol{\bchi}{\mathord}{bletters}{"1F}

\begin{document}

\def \om{\omega}
\def \vep{\varepsilon}
\def \vth{\vartheta}
\def \Dl{\Delta}
\def \dl{\delta}
\def \ph{\phi}
\def \ep{\epsilon}
\def \be{\beta}
\def \al{\alpha}
\def \si{\sigma}
\def \la{\lambda}
\def \iso{{\it iso}(3)}
\def \ISO{{\it ISO}(3)}
\def \ppu{\times\!\!\!\!\!\!\supset}
\def \pps{+\!\!\!\!\!\!\supset}
\def \L{{\cal L}_{\ph}}
\def \BR{I\!\!R}
\def \r{\vec r}
\def \cB{{\cal B}}
\def \cd{\partial}
\def \cds{\partial^{\,2}}
\def \D3{\Delta^{(3)}}
\def \pa{\ph^a_{\,\,\,\al}}
\def \da{\cd^2_{a a}}
\def \dc{\cd^2_{k k}}
\def \po{\cd^2_{1 1}}
\def \pm{\cd^2_{1 2}}
\def \p22{\cd^2_{2 2}}
\def \ddf{\frac{d}{\,d \rho\,}}
\def \lll{\matrix{\, \cr ^{\stackrel{+}{-}}\cr}}

\title{THE DISLOCATION STRESS FUNCTIONS\\
      FROM \\ THE DOUBLE CURL T(3)-GAUGE EQUATION:\\
       LINEARITY AND A LOOK BEYOND}
\author{{\bf C.~MALYSHEV}\\ \bigskip\\
V.A.Steklov Institute for Mathematics at St.Petersburg \\
Fontanka 27,~St.Petersburg 191011, Russia\\
 E-mail:~malyshev@pdmi.ras.ru}

\maketitle
\bigskip

\begin{abstract}
$T(3)$-gauge model of defects based on the gauge Lagrangian
quadratic in the gauge field strength is considered. The equilibrium
equation of the medium is fulfilled by the double curl Kr\"oner's
ansatz for stresses. The problem of replication of the static edge
dislocation along third axis is analysed under a special, though
conventional, choice of this ansatz.
The translational gauge equation is shown to constraint the
functions parametrizing the ansatz (the stress functions) so that the
resulting stress component $\sigma_{3 3}$ is not that of the edge
defect. Another translational gauge equation with the double curl
differential operator is shown to reproduce both the stress functions,
as well as the stress tensors, of the standard edge and screw
dislocations. Non-linear extension of the newly proposed translational
gauge equation is given to correct the linear defect solutions in next
orders. New gauge Lagrangian is suggested in the Hilbert--Einstein form.

\end{abstract}

\vskip 4truecm
\rightline{PDMI PREPRINT---3/1999}
\rightline{COND-MAT/9901316}

\newpage
\section{INTRODUCTION}

Considerable attention has been paid in last years to various
applications of the $\ISO$ ($\ISO$ is the group of rigid body
$T(3)\ppu SO(3)$) gauge model of defects in (continuous) solids
proposed in [1] and, as a more elaborated version, [2]. For instance,
the gauge theory of continuum damage in solids [3], the gauge theory
of elastic materials exhibiting a relaxation phenomenon [4], the gauge
theory of ``plastically incompressible'' elastic--plastic non-dissipative
medium [5] have been considered, as well as the problem of electronic
states in solids containing isolated defects [6, 7]. All these applications
[3--5, 7] have been mainly concerned with the translational sector of the
model [1, 2], i.e. with the Lagrangian $\L$ quadratic in the gauge field
strength postulated in [1, 2] to govern the translational gauge field.

The applications found look promising for the class of
the $T(3)$-gauge models with the quadratic Lagrangians like $\L$.
On the other hand, a conventional approach to the defects in solids,
i.e. the defect theory [8] which is concerned with the Volterra-type
solutions of incompatible elasticity theory, is widely acknowledged.
The theory [8] deals with the singular dislocations and disclinations,
as well as with their distributions, and it admits a number of reliable
calculations in the field of mechanical problems of solid state physics
[9] (for instance, and refs. therein). Besides, the use should be mentioned
of the singular solutions [8] in the problem of sound propagation in
elastic body with topological defects [10], and in such microscopic
problem as particles diffusion in solid with randomly distributed defects
[11]. For instance, the screw dislocation solution has been adapted both
in [10] and [11].

Therefore it is seen that the conventional approach [8], as well
as the non-conventional one [1, 2], are actively developed.
However, with regard to the defect modelling in the framework of the
Lagrangian $T(3)$-gauge approach, and namely with the quadratic $\L$,
the following problem lacks a sufficient attention by now:
whether some of the solutions considered in [8] can be reproduced
(`replicated', the terminology of [12]) in the formalism [1, 2]? The
Refs.[1, 2] themselves give a confirmative answer about this both for the
screw and edge dislocations (one should also refer to the original
paper [12]). Moreover, the last statement has stimulated the
attempt [7] to calculate the second order corrections to the screw
dislocation in the formalism [1, 2].

Conventional theory of defects [8] can be naturally considered as the
Abelian
gauge model [13--15] with the additive gauge group $\iso\approx\BR^6$
[16] ($\iso$ is the Lie algebra of $\ISO$). In this case the motor
[17] of the disclination and dislocation loop densities plays the role
of the 6-component Abelian gauge potential, and the motor of the
defect densities that of the Abelian gauge field strength [18]. It can
presumably be concluded from [18] that when $\ISO$ is gauged, a
successful replication (as soon as the last can be questioned) of the
defects [8] should be valid both for the translational and rotational
ones. Unfortunately, any attempt to obtain the standard wedge and
twist disclinations [8] in the formalism [1, 2] is seemingly absent.

The present paper has originated as an attempt to understand, whether
the Lagrangian $\L$ [1, 2] (i.e., practically, the translational model
[12]) leads to an extension of the classical picture [8] for (singular)
dislocations, or it provides a model of its own significance with the
solutions which should be interpreted on their own rights.

Formally, the double curl Kr\"oner's ansatz for stresses is used in [1, 2]
to fulfil the equilibrium equation of the medium. In its turn, the
translational gauge equation (the gauge equation, for brevity)
determines the stress functions which parametrize this ansatz since
can be re-expressed by means of the constitutive relation
`strain--stress'.
As to the screw defect, the gauge equation has been reduced
in [1, 2] to the homogeneous Helmholtz equation instead of the
standard Poisson equation for the stress function. In the axisymmetric
case ($\cd/\cd x_3\equiv 0$) the Helmholtz equation has provided the
modified Bessel function $K_0(\rho)$ as that solution, which coincides
with the Prandtl stress function [19], i.e. behaves as $\log(1/\rho)+
const$, at $\rho\ll 1$. Thus the screw defect is permissible in [1, 2].

In spite of the fact that the approach [12] to the screw dislocation is
not contradictive, the situation with the edge one is more subtle.
Generally speaking, the quadratic $\L$ becomes inappropriate as
soon as the stress functions of the edge dislocation (along third axis)
are obtained in the same way as for the screw one. Practically, the
$T(3)$-gauge equation constraints the appropriate stress functions so
that the resulting stress component $\sigma_{3 3}$ is not of the edge
defect.

An alternative form of the gauge equation is proposd in the present
work, which eliminates the contradiction and, without any artificial
effort, admits the linear solutions for the edge and screw dislocations.
It is crucial that both the gauge equations, i.e. the newly proposed
and that from [1, 2], being expressed through the general tensor of stress
functions, pass each into other under the exchange $\nu\longleftrightarrow
\nu^{-1}$, $\nu$ is the Poisson ratio. This is just the point needed to
avoid the problem found for $\si_{3 3}$, thought to keep the situation
with the screw dislocation unaffected.

The new gauge equation allows generalization, since can be
written as an Einstein-type (non-linear) equation. This
requires to modify the geometrical arena of the model [1, 2], and
suggests $\L$ in the Hilbert--Einstein form. The stress tensor
of non-linear elastic body plays the role of the source in the
Einstein-type gauge equation. As a specialization of this source, the
Murnaghan's constitutive relation of isotropic body is considered in the
present paper. The new gauge equation should be appropriate to find out
quadratic corrections to the linear solutions outside of the defect cores.

The paper is written in five sections. Sec.1 is introductory, Sec.2
is concerned with the difficulties of the replication of the static
edge dislocation, and a modified linear model (i.e. another
translational gauge equation, in fact) is proposed in
Sec.3. Further, Sec.4 is devoted to the Lagrangian formulation of the
corresponding non-linear model, and suggests
$\L$ in the Hilbert -- Einstein form, while Sec.5 concludes the paper.
In what follows, it is assumed that reader can be referred to [1, 2]
for motives of the $\ISO$-gauging, as well as for certain details
and comments on the original formulation. The consideration is time
independent, Greek indices run from 0 to 3, Latin ones -- from 1 to 3,
and repeated up and down indices mean summation. Besides, the Latin
letters a, b, c, d,...  denote curvilinear indices of a deformed
configuration, and i, j, k, l,... denote Cartesian frame of initial
(undeformed) one.

\section{T(3)-GAUGE EQUATION FROM THE QU\-AD\-RA\-TIC
LA\-GRAN\-GIAN AND RE\-L\-A\-T\-ED DIFFICULTIES}

Following [1, 2] let us briefly remind the basic elements of the theory
of the translational gauge field $\pa$ assuming that the gauge-rotational
degrees of freedom of the whole $\ISO$-model are ``frozen''. The
translational gauge field $\pa$ plays ``compensating'' role as the Abelian
group $T(3)$ acts by non-homogeneous shifts on the current configuration
variable $\xi^a$, and the corresponding compensated derivative has
the form $B^a_{\,\,\,\al}\equiv\cd_\al\xi^a + \pa$, where $\cd_{\al}$ are
the partial derivatives $\cd/\cd x_\al$, and  $x_0$ implies time.
The current configuration $\xi^a(\vec x, x_0)$ can be written in the
form $\xi^a = \dl^a_i\,x^i + u^a (\vec x, x_0)$ with respect to a Cartesian
coordinate system so that $\vec x$ implies an initial (undeformed)
configuration, and $\vec u$ is displacement field. It should be noticed
that when Abelian non-compact group like $T(3)\approx\BR^3$ is gauged,
appearance of $B^a_{\,\,\,\al}$ instead of the purely gradient quantity
$\cd_\al\xi^a$ implies that the latter ceases to be adequate variable,
because non-integrable contributions become essential.

As far as the problem of replication of the static edge dislocation
is in focus of the present paper (see [8] for the standard results),
it will be assumed that $\cd_0\equiv 0$,
$\ph^a_{\,\,\,0}\equiv 0$ so that the governing equations are:
$$
\cd_i\,\si^{\,\,\,i}_a\,=\,0\,,
\eqno(1.1)
$$
$$
\cd_j\,\left(\cd^j \ph^{\,\,\,i}_a\,-\,\cd^i\ph^{\,\,\,j}_a\right)\,=
\,\cd_j \cd^j \ph^{\,\,\,i}_a\,-\,\cd^i (\cd_j \ph^{\,\,\,j}_a)\,=
\,(2s)^{-1}\si^{\,\,\,i}_a\,.
\eqno(1.2)
$$
The source field $\si^{\,\,\,i}_a$ (1.2) is given by
$$
\si^{\,\,\,i}_a\,=\,B_{a j} \left( \la\,\dl^{i j}\,E^{\,\,\,k}_k\,+\,
                                  2 \mu\,E^{i j} \right)\,,
\eqno(2)
$$
$\la$ and $\mu$ are the Lam$\acute e$ constants, and the strain field
$E_{i j}$ is defined as $2 E_{i j}\,=\,B^k_{\,\,\,i}\,B_{k j}
\,-\,\dl_{i j}$ (see Sec.3 on the usage of up and down indices).
Notice that the stress field $\si^{i j}$ is just given by the brackets in
the R.H.S. of (2). It is seen that the equilibrium Eq. (1.1) ensures
integrability of (1.2). The parameter $s$ in the R.H.S. of (1.2) (the
``coupling'' constant) is due to the following choice of the spatial part
of the quadratic gauge-translational Lagrangian:
$$
\L\,=\,(-\,2s)\,\cd_{\,[ i}\ph^c_{\,\,\,j\,]}\,
              \cd^{\,[ i}\ph_c^{\,\,\,j\,]}\,,
\eqno(3)
$$
where the square brackets imply antisymmetrization. It has
been stated in [1, 2] that the translational gauge equations (1.2) ``are
what replace the compatibility conditions of linear elasticity theory''.

The Refs.[1, 2] propose the following approach to the dislocation
problem. The purely integrable contribution to the state due to $u_b$ is
zero. Therefore, Eq.(2) can be linearized in $\ph_{i b}$, since
$B_{i b}$ is $\dl_{i b}+\ph_{i b}$, as follows:
$$
\si_{i j}\,=\,2\mu\,\ph_{(i j)}\,+\,\la\dl_{i j}\ph_{k k}\,,
\eqno(4)
$$
where the brackets mean symmetrization $\ph_{(i j)} =$ $(1/2)(\ph_{i j}
+ \ph_{j i})$ (in linear case we do not distinguish up and down indices).
Further, we impose the conditions $\ph_{12}=\ph_{21}$,
$\ph_{13}=\ph_{31}=0$, $\ph_{23}=\ph_{32}=0$, and suppose $\cd_3\equiv0$
to adjust the axial orientation. In this case only a part of Eqs.(1.2)
survives:
$$
\begin{array} {cc}
\p22 \ph_{11}\,-\,\pm \ph_{12}\,=\,(2s)^{-1} \si_{1 1} & \\
\po \ph_{22}\,-\,\pm \ph_{12}\,=\,(2s)^{-1} \si_{2 2}\,, &
\end{array}
$$
$$
\eqno(5)
$$
$$
\begin{array} {cc}
\po \ph_{12}\,-\,\pm\ph_{11}\,=\,(2s)^{-1} \si_{12} &   \\
\p22 \ph_{12}\,-\,\pm\ph_{22}\,=\,(2s)^{-1}\si_{12} & ,
\end{array}
$$

$$
\Dl \ph_{33}\,=\,(2s)^{-1}\si_{3 3}\,,
\eqno(6)
$$
where $\Dl\equiv\po+\p22$. It can be realized that the choice of
$\ph_{i j}$ in the symmetric form $\ph_{i j} = \ph_{(i j)}$ should provide
us the appropriate solution of (1.1), (5), (6) as a small elastic strain
tensor, while Eq.(4) gets the status
of the constitutive relation of isotropic body (the Hooke law).

The way to handle the system (1.1), (5), (6) proposed in [12] is as follows:
to postulate the ansatz
$$
\mu^{-1}\bsigma\,=\,\left(
\begin{array} {ccc}
\,\,\,\p22 f  &      -\pm f & 0 \\
      -\pm f  & \,\,\,\po f & 0 \\
\,\,\,\,\,\,0 & \,\,\,\,\,0 & p
\end{array}\right)\,,
\eqno(7)
$$
which fulfils (1.1), while Eqs.(5), (6) enable to specify
the unknown functions $f=f(x_1,x_2)$ and $p=p(x_1,x_2)$ provided the use
of the linear law (4) is made.

As far as we are interested in the edge dislocation along the third axis
$Ox_3$, the component $\ph_{3 3}$ (i.e. the strain $(3 3)$-component)
is expected to become zero, at least as a limit of an ``extended''
solution. As $\si_{3 3}=\nu(\si_{1 1}+\si_{2 2})$, where $\nu=\la/2
(\la+\mu)$ is the Poisson ratio, implies vanishing of the strain
$(3 3)$-component, the constraint $\nu\Dl f=p$ has to be expected in that
limit.
The relation (7) with $p=\nu\Dl f$ but, formally, with opposite sign at
$f$ is nothing but completely the standard ansatz of the theory of
dislocations, where $f$ is called Airy's stress function [19]. This
function $f$ has been found in [19] as the biharmonic potential
$\cd_2(\rho^2\log\rho)$ (or $\cd_1(\rho^2\log\rho)$),
$\rho^2=x^2_1+x^2_2$, and it enables to obtain
via (7) all the stress tensor components of the edge defect.

Let us re-consider the solution [1, 2] of the problem of the edge
dislocation. Clearly, Eq.(1.1) is respected by
{\mathversion{bold}$\si$} (7), whereas Eqs.(5) result in
$$
(1-a)\Dl f\,-\,a\,p\,=\,\kappa^2 f\,,
\eqno(8)
$$
and (6) leads to
$$
(1-a)\Dl p\,-\,a\,\Dl\,\Dl f\,=\,\kappa^2 p\,,
\eqno(9)
$$
where $a=\la(3\la+2\mu)^{-1}$ and $\kappa^2\equiv\mu/s$. Let us
exclude $p$ from (9) with the help of (8). We assume here the limit
$\kappa\to 0$ proposed in [12] to restore the classical situation [8].
In this case $\Dl \Dl f=0$ arises to define the limiting form of $f$,
while (8) provides
$$ p\,=\,\frac{1-a}{a}\,\Dl f\,\equiv\,
\frac{\,1\,}{\nu}\,\Dl f \eqno(10) $$
as the limiting form of $p$.

The point is that the fourth order differential equation, which appears to
govern $f$ at arbitrary $\kappa$, looks troublesome for analytical
solution. It is why, in view of $\Dl\Dl f=0$, the authors have been
forced to ``guess'' that $f$ can be found so that the needed Airy's
function behaviour appears for it at $\kappa\to 0$. But since
$\Dl\Dl f=0$ is to indicate at the correct $f$, just the same reasons
(i.e. linearity of the substitution of $p$) imply that Eq.(10)
defines the limiting form of $p$. However, the Eq.(10) contradicts the
constraint demonstrated above. Thus, in spite of the satisfactory
equation for $f$, the correct stress component $\si_{3 3}$ of
the edge dislocation would not appear neither at finite $\kappa$ nor at
$\kappa\to 0$.  Notice that in the Ref.[1] it has been recognized only
for finite $\kappa$ that the new $\si_{3 3}$ is different from the
standard one.

Otherwise, let us try ``rigidity'' of (8), (9) against the constraint
$p=\nu\Dl f$ by substituting it to them:
$$
(1-\nu)\,\Dl f\,=\,\kappa^2 f\,,
$$
$$
0\,=\, \kappa^2 \Dl f\,.
$$
These equations can hardly be fulfilled with nontrivial $f$
at finite $\kappa$, and, moreover, nontrivial solutions of $\Dl\Dl f=0$
with the analytical behaviour we are interested in would not appear
even at $\kappa\to 0$.

Therefore, the joint use of Eqs.(5), (6) and (7) to obtain the
edge dislocation's stress function leads to the following conclusion.
The resulting Eqs.(8) and (9) constraint the stress functions $f$ and
$p$ so that it is impossible to get $f$
as the biharmonic potential, and to fulfil at the same time $p =\nu\Dl
f$. One gets either only $\si_{3 3}$ is incorrect due to (10), or $f$ is
not the Airy function at all. Besides, there is a little hope that an
improved $f$ can be found at finite $\kappa$ so that the additional
contribution $(-1/a) \lim_{\kappa\to 0} \left(\kappa^2 f \right)$
from (8) to the R.H.S. of (10) could restore the desired $p=\nu\Dl f$.

With the purpose of illustration, let us obtain the limiting form of the
matrix {\mathversion{bold}$\ph$} by means of (4) inverted and (7). We shall
substitute the known $f=(-C/2) \cd_2(\rho^2\log\rho)$, where $C=(b/2\pi)(1-
\nu)^{-1}$, but $p$ will be evaluated accordingly to (10). Then we get:
$$
\bphi\,=\,\frac{\,C\,}{2\,\rho^2}
\left(\begin{array} {ccc}
x_2\left(1-2\frac{x^2_1}{\rho^2}\right) &
x_1\left(1-2\frac{x^2_2}{\rho^2}\right) & 0 \\
x_1\left(1-2\frac{x^2_2}{\rho^2}\right) &
x_2\left(1+2\frac{x^2_1}{\rho^2}\right) & 0 \\
\,\,\,\,\,0 & \,\,\,\,0 & 2(1-\frac{\,1\,}{\nu})\,x_2
\end{array}\right)\,.
\eqno(11)
$$
Integrating the equations $\cd_1 u=\ph_{1 1}$, $\cd_2 v=\ph_{2 2}$,
$\cd_2 u+\cd_1 v=2\ph_{1 2}$ with respect to the new fields $u, v$,
we obtain:
$$
v\,+\,i\,u\,=\,v_d\,+\,i\,u_d\,+\,\frac{\,b\,}{2\pi}\log(x_1\,-\,i\,x_2)\,,
\eqno(12)
$$
where $i=\sqrt{-1\,}$, and $u_d$, $v_d$ are the edge dislocation's
displacements in the plane perpendicular to $Ox_3$ (the Burgers vector is
in $x_1$-direction) [8]. Equation (12) is to express that namely the
logarithmic term responsible for the ``closure failure'' just
characterizing the defect does not enter to $u, v$, because $v_d+i u_d$
contains $(-b/2\pi)\log(x_1-i x_2)$. Besides, $\ph_{3 3}$ (11) looks
embarrassantly to consider it as the corresponding strain component of
the plane problem in question. In spite of the statement by [1, 2],
it is plausible to conclude that Eq.(11), as the
limiting form of possible solution to (5), (6), does not imply
at all replication of the edge dislocation at $\kappa\to 0$.

It can occur that additional care with $\kappa \to 0$ is needed
to make rigorous statement about absence (or presence) of the edge
dislocation solution for the Eqs.(5), (6). The formal implications of
the use of Eqs.(4), (5)--(7) have only been pointed out here. In
the next section a modification of the situation, i.e. a way to keep
the ansatz (7), to exchange the parameters $a$ and $1-a$, and thus to
avoid the unpleasant $p=\nu^{-1}\Dl f$, will be demonstrated which
admits the standard dislocations in the formal limit $\kappa\to 0$.

\section{THE DOUBLE CURL T(3)-GAUGE EQUATION}

The equilibrium equation (1.1) can be satisfied identically provided
the stress field {\mathversion{bold}$\si$} is chosen as double curl
of a twice differentiable symmetric tensor potential $\bchi$ which is
called [19] the tensor field of second order stress functions:
$$
\si_{i j}\,=\,({\rm inc}\,\bchi)_{i j}\,
\equiv\,-\,\ep^{ikl}\,\ep^{jmn}\, \cd^2_{k m}\chi_{l n}\,.
\eqno(13)
$$
Both the particular ansatz proposed in [1, 2] to obtain the edge
and screw dislocations appear as specializations of the general
representation (13) (though (13) itself is not stressed in [1, 2]).

The so-called stress function method based on (13) has been developed in
[20, 21] to approach to non-linear dislocation problems. For the
internal stress problem (isotropic case) the representation (13) has also
been discussed by
Kr\"oner in [19]. The concept of a torsion-free stress space and
relationship to it of the Eq.(13) have been considered in [19]. In [22]
a stress--strain duality has been presented which relates
the double curl ansatz (13) to a similiarity in Riemannian descriptions
of kinematics and statics of mechanical state of solid. It has been
exploited in [22] that the tensor of secon order stress functions and
the stress tensor could play the role of the metric and the Einstein
tensors, accordingly, of a non-Euclidean stress space.

Therefore Eq.(7) looks more fundamentally than Eqs.(5), (6) with
regard to the unsatisfactory implications of $p$ (10). On the other
hand, it is just the choice of the master equation (1.2) for the
translational field which would require a justification.
It turns out that the most direct way to get the function $p$
consistent with the requirement $\ph_{3 3}=0$, i.e. $\si_{3 3} =
\nu(\si_{1 1} + \si_{2 2})$, at $\kappa\to 0$, is to
choose the L.H.S. of Eq.(1.2) (linear consideration) as follows:
$$
({\rm inc}\,\bphi^{Sym} )_{i j}\,\equiv
\,-\,\ep^{ikl}\,\ep^{jmn}\,\cd^2_{k m} \ph_{(l n)}\,=
\,(2s)^{-1}\si_{i j}\,,
\eqno(14)
$$
where superscript $^{Sym}$ implies tensor symmetrized, $\ep^{ikl}$ is the
permutation symbol, and $\si_{i j}$ is given by (4). When the R.H.S. of (14)
is zero, it looks like a compatibility equation of linearized elasticity
theory provided $\ph_{i j}$ is considered as distortion. Equation (14)
replaces the compatibility condition in the sense that its R.H.S. is
non-trivial while the L.H.S. looks conventionally. In the next section
Eq.(14) will be viewed as the Einstein-type one.

Let us accept (14) instead of the linearized (1.2) in the $T(3)$-gauge
model of defects. However, it is more appropriate to rewrite (14) through
the unknown second order stress functions $\bchi$. To this end one should
express $\bphi^{Sym}$ using both the Hooke law (4) inverted and (13), and,
after this, substitute it into the L.H.S. of (14).
The stress field {\mathversion{bold}$\si$} (13) takes its place in the
R.H.S. of (14). Finally, the fourth order differential equation appears:
$$
\D3 \D3 \chi_{i j}\,+\,a\,{\rm D}_{i j}\D3\chi\,+
\,\left((1-a)\,\cds_{i j}\,+\,a\,\dl_{i j} \D3 \right)
\cds_{k l} \chi_{k l}\,- \,\D3\left(\cds_{i k} \chi_{j k}\, +\,\cds_{j
k} \chi_{i k}\right)\,=
$$
$$
=\, \kappa^2 \left(\D3 \chi_{i j}\,+\,{\rm D}_{i j} \chi\,+\, \dl_{i j}
\cds_{k l} \chi_{k l}\,-\,\cds_{i k} \chi_{j k} \,-\,\cds_{j k}
\chi_{i k}\right)\,,
\eqno(15)
$$
where $\chi$ implies trace of $\bchi$, $a$ is defined by Eq.(10),
$\kappa$ is defined in (8), (9), and
we have denoted the differential operators $\D3 = \dl_{i j}\,\cds_{i j}$,
and ${\rm D}_{i j} = \cds_{i j} - \dl_{i j}\,\D3$. In its turn, the
Eq.(15) can be considerably simplified if we replace $\bchi$ by
another symmetric potential $\bchi^{\prime}$ as follows:
$$ \chi_{i j}\,=\,2 \mu\left(\chi^{\prime}_{i j}\,+\,\frac{\,\nu\,}
{1-\nu}\,\dl_{i j} \chi^{\prime} \right)\,,
$$
where $\bchi^{\prime}$ fulfils $\cd_i\chi^{\prime}_{i j}=0$. Thus we get:
$$
\D3 \D3 \chi^{\prime}_{i j}\,=\,\kappa^2 \left(\D3 \chi^{\prime}_{i j}
\,+\,\frac{\,1\,}{1-\nu} {\rm D}_{i j} \chi^{\prime} \right)\,.
$$

To clarify the situation let us repeat the procedure leading to (15) for
the Eq.(1.2) as well, though assuming that the latter is written in terms
of $\bphi^{Sym}$. It is curious that thus resulting equation is completely
similiar to (15) except only one thing: $a$ and $1-a$ are just exchanged.
In other terms, $\nu$ and $\nu^{-1}$ are exchanged. Therefore it can be
guessed that the problem related
to $p$ (10) should disappear when (14) is used instead of (1.2).

It has to be noticed that linearizing Eq.(1.2) one gets another,
apart from the problem of $p$, unpleasant thing: the source $\si_{i j}$
(4) in its R.H.S. is symmetric in the indices, while the L.H.S. is
obviously not. It is why in [1, 2] $\ph_{i j}$ has been additionally
assumed symmetric to perform the concrete calculations. On the contrary,
the L.H.S. in (14) is symmetric, and the antisymmetric part of the
translational gauge field automatically drops out, but must be taken into
account to restore, say, distortion from strain field.

In order to specialize (15) to the edge dislocation case, we
introduce two functions $p$ and $f$ as follows:
$$
f\,\equiv\,\chi_{3 3}\,, \qquad p\,\equiv\,-\,\cds_{2 2}
\chi_{1 1}\,-\,\cds_{1 1} \chi_{2 2}\,+\, 2 \cds_{1 2} \chi_{1 2}\,,
$$
while other $\chi_{i j}$ are zero. The plane problem is adjusted
by $\cd_3\equiv 0$, and thus $\D3$ becomes $\Dl$ (see Sec.2 for $\Dl$).
It can be verified that now six equations (15) are reduced to
$$
a\Dl f\,+\,(1-a)\,p\,=\,\kappa^2 f\,,
\eqno(16)
$$
$$
(1-a)\Dl\Dl f\,+\,a\,\Dl p\,=\,\kappa^2 p\,.  \eqno(17)
$$
More precisely, Eq.(16) appears instead of the three equations
$\cds_{1 1} A=0$, $\cds_{1 2} A=0$, and $\cds_{2 2} A=0$ where
$A\equiv (1-a) p+ a \Dl f-\kappa^2 f$. It is seen that changing $f$ to
$-f$ and $a\longleftrightarrow 1- a$ one obtains (8), (9) from (16),
(17), accordingly.

Therefore Eq.(14) has provided us with the remarkable opportunity to
convert the embarrassing ratio $(1-a)/a$ in (10) into $a/(1-a)$.
Finally, the resulting function $p$ (16) gets around the obstacle
discussed in the previous section. Notice that choosing $f\equiv
-\cd_{1} \chi_{2 3} +\cd_{2} \chi_{3 1}$ one can deduce from (15) a
single equation
$$ \Dl f\,=\,\kappa^2 f\,, \eqno (18) $$
which is the same as in [1, 2] for the screw dislocation ansatz.
It is clear that this coincidence is because (18) does not contain
the Poisson ratio $\nu=a/(1-a)$. Solutions to (16)--(18)
(and, generally, to (15) ) should
be called ``modified stress functions'' to distinguish them from
the classical harmonic and biharmonic potentials.

As the Eq.(16) defines $p$, the second equation governing $f$ gets the form:
$$
\left(\Dl-{\cal M}^2 \right)\,\left(\Dl+ {\cal N}^2 \right) f\,=\,0\,,
\qquad {\cal N}^2 \equiv \frac{\,\mu\,}s \frac{\,1\,}{\,1-2a\,}\,.
\eqno(19)
$$
For correspondence with [1] we use ${\cal M}^2$ instead of
our $\kappa^2$ when considering (19) and its solution.
Equation (19) simply differs from the corresponding Eq.(4.6.27) in [1]:
only $2a-1$ and $1-2a$ are interchanged under $a \longleftrightarrow
1-a$.

As $\cd_2$ commutes with $\Dl$, one can be concerned with $f$ in the
form $\cd_2 h (\rho)$, $\rho=|\vec x|$, so that $h$ also respects
(19) with $\Dl$ reduced to
$$
\frac1{\,\rho\,} \ddf \left(\rho \ddf \right)\,.
$$
Equation (19) has the Bessel and Neumann functions
$J_0({\cal N}\rho)$ and $Y_0({\cal N}\rho)$, and the modified Bessel
functions $I_0({\cal M}\rho)$, $K_0({\cal M}\rho)$ as four
angle-independent basic solutions [23].  These basic solutions can be
combined so that their combination is decreased at infinity.
Eventually, the following solution can be
obtained:
$$
f(\vec x)\,=\,\frac{\,b s\,}{2 \pi}\,\cd_2\,{\cal
F}(\rho)\,,
$$
$$
\eqno(20)
$$
$$
{\cal F}(\rho)\,=\,\log\frac{{\cal N}}{{\cal M}}\,
J_0({\cal N}\rho)\,-\,\frac{\pi}2\, Y_0({\cal N}\rho)\,-\,
K_0({\cal M} \rho)\,.
$$
At ${\cal M}\rho, {\cal N}\rho \ll 1$, the solution $\,f$ (20) results
in the standard biharmonic potential, since
${\cal F}\simeq {\cal M}^2/ (2(1-\nu)) \rho^2\log\rho$. Notice, that
the normalization of ${\cal F}(\rho)$ is chosen so that the stresses
are $\si_{1 1} = -\cd^2_{2 2} f$, etc., but not, accordingly to (7),
$\si_{1 1} = -\mu\,\cd^2_{2 2} f$, etc..  The limit ${\cal M}\rho,
{\cal N}\rho \ll 1$ implies that $\rho$ is finite but not zero while
$\mu /s\to 0$. In the opposite case
${\cal M}\rho, {\cal N}\rho \gg 1$, the solution (20) becomes a linear
combination of $x_2\,\rho^{-3/2} \sin({\cal N}\rho)$ and
$x_2\,\rho^{-3/2} \cos({\cal N}\rho)$, i.e. it is
$O\,\left(({\cal N}\rho)^{-1/2}\right)$. Therefore, we are expecting
all the entries of the stress matrix {\mathversion{bold}$\si$} at
${\cal M}\rho, {\cal N}\rho \ll 1$ to be those of the edge
dislocation, because $p=\nu\Dl (-f)$ by $\kappa\to 0$ in (16).
Recall that the solution to (18) $K_0$, which enables the screw
dislocation, possesses fast and monotonic decreasing as $\rho^{-1/2}
\exp({-\kappa}\rho)$. In other words, this solution is really
`short-ranged' in comparison with the corresponding classical one.

Let us also mention the Ref.[39], where a ``mass'' of defects is
discussed as an implication of the translational model [1] (specifically,
as implication of the corresponding time-dependent $T(3)$-gauge
equation). This is because the Klein--Gordon equation is possible with
the mass $\kappa^2$ for the translational gauge field in the
Lorentz-like gauge. The present section demonstrates, that such
``mass'' effect should be rather traced to Eq.(15) for the
modified stress functions. The point is that namely (14), but not
(1.2), seems to be related to the defects (at least, to the statics of
the conventional ones). Moreover, it is just the Eq.(15), which
implies the conventional, unboundly increasing, stress functions to
become the limited modified ones just due to the ``mass terms''
$\kappa^2$ in (18) and (19).

However, in Sec.4.3, and 5, it will be
discussed, that only the replication demonstrated above
(i.e. the regime $\kappa\rho \ll 1$) is of main interest
in the linearized version of the Einstein--type consideration.
Thus, the L.H.S. of (14) helps to avoid the
problem with $p$ (10). Both the sides of (14) admit non-linear
generalizations, and therefore it is suggestive to put (14) into a more
general form which can be derived in Lagrangian approach.

\section{THE EINSTEIN - TYPE EQUATION AND ITS LAGRANGIAN
DERIVATION}

\subsection{The geometric preliminaries}

Before to proceed with the Lagrangian derivation of Eq.(14),
let us briefly present the geometrical apparatus which will underly
our main construction. To make the discussion reasonably compact, it
will be assumed that one can be referred, say, to [24] for basics of
geometry of the Riemann and Riemann--Cartan spaces in the
form accomodated to describe defects. Other useful references can also
be found in [25]. As to gauging of the important for us group $\ISO$,
a formally close gauging of the Poincar$\acute e$ group (which also
is a semi-direct product of translation and pseudo-orthogonal rotation
groups though of 4-dimensional Minkowskian space-time) have already been
extensively developed in the realm of gravitational physics [26]. One
should consult with [26, 27] for a similiar, though
much more elaborated machinery.

Although we are restricted to the $T(3)$-gauging, it is more appropriate
to admit, just for a moment, a more general framework of the
$\ISO$-gauging. Here the couple of the Cartan structure equations
\def \cR{{\cal R}^{a}_{\,\,\, b,\,i j}}
\def \vt{{\cal T}^a_{\,\,\,,i j}}
\def \Gm{\Gamma}
$$
\cR\,=\,\cd_i\,\om^{a}_{\,\,\, b,\,j}\,-\,
        \cd_j\,\om^{a}_{\,\,\, b,\,i}\,+
  \om^a_{\,\,\,c,\,i}\,\om^{c}_{\,\,\, b,\,j}\,-\,
  \om^a_{\,\,\,c,\,j}\,\om^{c}_{\,\,\, b,\,i}\,,
$$
$$
\eqno(21)
$$
$$
\vt\,=\,\cd_i B^a_{\,\,\,j}\,-\,
 \cd_j B^a_{\,\,\,i}\,+\,\om^a_{\,\,\,c,\,i}\,B^c_{\,\,\,j}
 \,-\,\om^a_{\,\,\,c,\,j}\,B^c_{\,\,\,i}\,.
$$
appears as one of the basic differential--geometric relations.
Four tensors which enter the Eqs.(21) are components of the
differential forms which define the geometry of the so-called
Riemann--Cartan spaces: $\cR$ and $\vt$, which are antisymmetric in
$i, j$, determine the curvature and torsion 2-forms, accordingly
${{\cal R}^a}_b$ and ${\cal T}^a$.  The coefficients $\om^{a}_{\,\,\,
b,\,i}$ and $B^a_{\,\,\,i}$ define the connection 1-form ${\om^a}_b$
and the co-frame 1-form $B^a$, respectively.  The couple of indices
$a, b$ demonstrate ${{\cal R}^a}_b$, ${\om^a}_b$ as elements of the
Lie algebra of the group $SO(3)$, while ${\cal T}^a$, $B^a$ belong to
the Lie algebra of the group $T(3)$. All the fields are considered in
open domain of a 3-dimensional manifold with Euclidean signature.

It is known that gauging $\ISO$ one gets the $\iso$-valued
connection and curvature differential forms, both of which are split into
``linear'' and ``translational'' parts owing to the
semi-direct sum structure of $\iso$ [27, 28]. However, the ``translational''
curvature is transformed non-covariantly under the $\ISO$ gauge
transformation. In order to find out the quantity which is transformed
(gauge-)covariantly, it is proposed to use the auxiliary field $\xi^a$
[27--29]. As the result, the Cartan structure equations in the $\ISO$
gauging acquire the form (21) provided the coefficients $B^a_{\,\,\,i}$
have the following structure:
$$
B^a_{\,\,\,i}\,=\,\ph^a_{\,\,\,i}\,+\,
\cd_i\,\xi^a\,+\,\om^a_{\,\,\,b,\,i}\xi^b\,.
\eqno(22)
$$
In (22) $\om^a_{\,\,\, b,\,i}$ and $\ph^a_{\,\,\,i}$ are just the
``linear'' and ``translational'' parts of the $\iso$-valued connection
1-form.  In other terminology, $\om^a_{\,\,\, b,\,i}$ and $\ph^a_{\,\,\,i}$
are the gauge potentials which are due to non-homogeneous action of $\ISO$.
Going further, apart from $B^a_{\,\,\,i}$, it is also appropriate to define
their reciprocals $\cB^{\,\,\,j}_b$ as follows:
$B^a_{\,\,\,i}\,\cB^{\,\,\,i}_b\,=\,\dl^a_b$,
$B^a_{\,\,\,i}\,\cB^{\,\,\,j}_a\,=\,\dl^j_i$.
The point is that the components $\cB^{\,\,\,j}_b$ define an
orthonormal triad $\cB_b = \cB^{\,\,\,j}_b\cd_j$ (more exactly,
$\cB^{\,\,\,j}_b$ are transitions between
the coordinate basis in the tangent space $\{ \cd_j \}$ and the
orthonormal basis $\{ \cB_b \}$), whereas $B^a_{\,\,\,i}$ provide the dual
basis of 1-forms $B^a = B^a_{\,\,\,i} d x^i$.

It is crucial that the auxiliary field $\xi^a$ (22) has been identified
in [1, 2] as the deformed configuration variable (see [28] for other
problem-motivated interpretations of that special field) as follows:
$$
\xi\,:\,\,x^i\,\longrightarrow\,\xi^a(x^i)\,\,,
\qquad \xi^{-1}\,:\,\,\xi^a\,\longrightarrow\,x^i(\xi^a)\,.
\eqno(23)
$$
This equally means that the group indices $a, b, c$ in (21) get the
``material'' meaning since the sets $\{\xi^a\}$ label points
in deformed configuration. The coordinate indices $i, j, k$ in (21)
get the status of the Cartesian ones in initial (undeformed) configuration.
In particular, both $B^a_{\,\,\,i}$ and $\cB^{\,\,\,i}_a$ are useful
to pass from one set of indices to another.

In [18] the $\ISO$ Cartan's equations (21) have been reduced to the
conventional expression of the motor of dislocation and disclination
densities through the motor of the defect loop densities. It is essential
that the field $\xi^a$ which is $x^a + u^a$ has enabled this truncation
so that sequence of Schaefer's exterior differentiations inherent to [8]
has naturally appeared. Besides, interpretation of the $\ISO$-connection
in terms of the defect loop densities has become possible.

Regarding the material interpretation of $\{x^i\}$ and $\{\xi^a\}$, let
us recall how the Green deformation tensor $g_{i j}$ and the corresponding
Lagrangian strain tensor $E_{i j}$ appear in the conventional elasticity
theory [30--32]. Indeed, the length element in a deformed configuration
can be written in the Euclidean form $ds^2\, =\, \eta_{a b} d\xi^a d\xi^b$
($\eta_{a b}$ is flat metric) as well. At the same time, the configuration
$\{\xi^a\}$ can be considered in terms of the initial one by means of
the maps (23). In this case the length element becomes expressed
through the initial coordinates:$ds^2 = \eta_{a b} \cd_i \xi^a
\cd_j\xi^b d x^i d x^j$. Thus we obtain the corresponding
non-Euclidean metric (the Green deformation tensor) $g_{i j}=\eta_{a
b}\,\cd_i \xi^a \cd_j\xi^b$, while the Lagrangian strain tensor is
$$
2 E_{i j}\,=\,g_{i j}\,-\,\eta_{a b} \dl^a_i \dl^b_j \,=\,
g_{i j}\,-\,\eta_{i j}\,.
\eqno(24)
$$
Following the close analogy, we define the Green deformation
tensor as
$$
g_{i j}\,=\,\eta_{a b}\,B^a_{\,\,\,i}\,B^b_{\,\,\,j}\,,
\eqno(25)
$$
with $B^a_{\,\,\,i}$ (22), while the strain tensor is given by (24).
The metric $g_{i j}$ and its inverse can be used for raising and
lowering the indices $i, j$, while $\eta_{a b}$ -- to handle
analogously $a, b, c$.

Now it is time to stress that the present work is concerned with the
Eq.(14) to avoid the problem discussed in the Sec.2. The given section
is to extend straightforwardly (14) as an
Einstein-type equation in the context of torsion-free Riemannian
geometry. Therefore, given the metric $g_{i j}$ is (25), a unique
linear connection without torsion $\Gm^k_{i j}$ can be associated with
it so that $g_{i j}$ is covariantly constant.
The corresponding condition $\nabla_i\,g_{j k}\,=\,0$ implies the
conventional expression for such $\Gm^k_{i j}$:
$$
2\,\Gm^k_{i j}\,=\,g^{k l}\, \left(\cd_i g_{j l}\,+\,\cd_j g_{i l}
\,-\,\cd_l g_{i j} \right)\,.
$$
The connection $\Gm^k_{i j}$ can be related with the $SO(3)$-connection
$\om^a_{\,\,\,b,\,i}$ by means of the relation
$$
\om^a_{\,\,\,b,\,i}\,=\,\cB^{\,\,\,l}_b\,(\Gm^k_{i l}\,B^a_{\,\,\,k}
\,-\,\cd_i B^a_{\,\,\,l})\,,
$$
which respects the covariant constance of $g_{i j}$, and, further,
allows to obtain from $\cR$ (22) the Riemann--Christoffel curvature tensor
${\rm R}^k_{\,\,\,m i j}$:
$$
\cR\,=\,{{\cal B}_b}^m \,B^a_{\,\,\,k}\,{\rm R}^k_{\,\,\,m i j}\,,
$$
$$
{\rm R}^k_{\,\,\,m i j}\,=\,
\cd_i \Gm^k_{j m}\,-\,\cd_j \Gm^k_{i m}\,+\,
\Gm^k_{i n}\,\Gm^n_{j m}\,-\,\Gm^k_{j n}\,\Gm^n_{i m}\,.
$$
As the last step, we define the scalar curvature ${\rm R}$,
which is an important geometric invariant as follows:
${\rm R} = {\rm R}_i^{\,\,\,i}$, ${\rm R}_{i j} = {\rm R}^k_{\,\,\,i k j}$.
The scalar ${\rm R}$ is just what we need to derive (14) in the Lagrangian
approach. From now on, we ``switch off'' the gauge--rotational degree of
freedom $\om^{a b}_{\,\,\,\,,i}$ in $B^a_{\,\,\,\,i}$.

The geometric presentation above is inevitably sketchy, since many
appropriate things,
like definition of principle bundle of affine (linear) frames, definition
of associated vector bundle, gauge transformation rules, etc., are omitted.
Necessary matter can be picked up from the literature cited, though more
detailed and accurate geometric background of the non-Abelian model would
require a separate presentation. However, as to the $\ISO$ gauge model
[1, 2], the idea to pass from the ``triad'' representation (21) to the
purely ``coordinate'' one, in order to discuss replication of the
conventional dislocations, belongs, seemingly, to this paper.

\subsection{The Lagrangian derivation}

It is well known that the six compatibility equations of elasticity
are, in fact, vanishing conditions of the six independent components of
the 3-dimensional
Riemann -- Christoffel tensor ${{\rm R}^i}_{j k l}$. Due to
3-dimensionality, one can equally use the second rank Einstein tensor
${\rm G}^{i j}$ instead of ${{\rm R}^i}_{j k l}$ as follows: ${\rm G}^{i j}
= (1/4) e^{ikl} e^{jmn} {\rm R}_{klmn}$, where
$e^{ijk}=\sqrt{g\,}\ep^{ijk}$ and $g$ is $det(g_{i j})$. Linearizing the
Riemann--Christoffel tensor one gets the double curl expression like the
L.H.S. of (14).

Therefore, the Eq.(14) can be considered as a weak
field approximation of an Einstein-type equation. Indeed, the L.H.S. of
(14) looks like linearization of ${\rm G}^{i j}$, while the stress tensor
in its R.H.S. can acquire higher powers of $E_{i j}$, as a non-linear
constitutive relation `stress--strain'. In
other words, the situation is reminiscent to a gravitational equation
which relates the Einstein tensor (though 4-dimensional) to a matter
energy--momentum tensor as the source. Therefore both the sides of (14)
can be extended, and the resulting equation admits a Lagrangian derivation.

Now we can proceed with the Lagrangian approach.
First of all, we postulate the Hilbert--Einstein Lagrangian density
which is responsible for the translational field:
$$
\frac{\,1\,}b\,\L\,=\,s\,{\rm R}\,,
$$
where $b \equiv det(B^a_{\,\,\,i})$, and ${\rm R}$ is the scalar curvature.
Variation of $\L$ takes the
form (up to terms irrelevant by the Stokes theorem):
$$
\frac{\,1\,}b\,\frac{\,\dl \L\,}{\dl B^a_{\,\,\,i}}\, =\,2 s\,\left(
{\rm R}_a^{\,\,\,i}\,-\,\frac{\,{\rm R}\,}2\,
{\cal B}_a^{\,\,\,i}\right)\, \equiv\,2 s\,{\rm G}_a^{\,\,\,i}\,,
\eqno(26)
$$
where ${\rm G}_a^{\,\,\,\,i}$ is the Einstein tensor.

Let us obtain {\mathversion{bold}$\si$} in the R.H.S. of the Einstein-type
equation. The field $\xi^a$ is an important constituent of more general
$\ISO$-formalism, and it should be governed by an appropriate Lagrangian
${\cal L}_\xi$. Our consideration is static, and therefore we shall choose
$(-1/b){\cal L}_\xi$ in the form of potential energy $W$ of isotropic
non-linear elastic continuum (practically, in the so-called Murnaghan's
form). Given the field $B^a_{\,\,\,i}$ has the $T(3)$-invariant
form $\cd_i\xi^a+\ph^a_{\,\,\,i}$, variation of the Lagrangian
${\cal L}_\xi$ gets the form:
$$
\frac{\,1\,}b\,\frac{\,\dl{\cal L}_\xi\,}{\dl B^a_{\,\,\,i}}\,=\,
-\,\Sigma^{\,\,\,i}_a\,\equiv\,-\,\si_a^{\,\,\,i}
\,-\,{\cal B}_a^{\,\,\,i}\,W\,.
\eqno(27)
$$
In (27), by definition, the stress field $\si^{i j}$ is given by
$\dl W/\dl E_{i j} = \si^{i j}$, and the second term in
$\Sigma_a^{\,\,\,i}$ is due to variation of $b$.

Putting together (26) and (27), we obtain the equation
$$
{\rm G}_a^{\,\,\,i}\,=\,(2 s)^{-1}\,\Sigma_a^{\,\,\,i}\,,
\eqno(28)
$$
which generalizes (14) in the sense explained above. Rejecting in (28)
higher terms and assuming coincidence, in the leading order, of its source
with (4) (weak field approximation), we obtain (14). Tending $s$ to
infinity, we just recover the general compatibility equation
${\rm G}_{i j}=0$. As far as the Einstein tensor is ``covariantly
conserved'' by the appropriate Bianchi identity [24], the equation
$\nabla_i \Sigma_a^{\,\,\,i} = 0$ appears to govern the source tensor.
For the first time in the context of gauge
dislocations, an equation similiar to (28) has been obtained by
variational approach in [33], but without any further elaboration.

For definitness, let us specialize the potential energy $W$ taking it
in the form proposed by Murnaghan [34]:
\def \ie{I\!\!E}
$$
W\,=\,\frac{\,\la\,}2 I_1^2(\ie)\,+\,\mu\,I_1(\ie^2)\,+
\,\frac{\,\nu_1\,}6 I_1^3(\ie)\,+\,\nu_2\,I_1(\ie)\,I_1(\ie^2)
\,+\,\frac{\,4\,}3\,\nu_3\,I_1(\ie^3)\,,
\eqno(29)
$$
where $\ie$ implies the gauge invariant strain tensor $E_{i j}$ (24),
invariant function $I_1 (...)$ implies trace of the appropriate tensor
argument, while $\la$ and $\mu$ are the Lam$\acute e$ constants of second,
and
$\nu_{1,\,2,\,3}$ -- of third orders. The stress tensor acquires the form
$$
\si^{i j}\,=\,\dl^{i j}\,
\left[\la\,I_1(\ie)\,+\,\frac{\,\nu_1\,}2\,I_1^2(\ie)
\,+\,\nu_2\,I_1(\ie^2)\right]\,+
\,2\,E^{i j}\left[\mu\,+\,\nu_2\,I_1(\ie)\right]\,+\,
4 \nu_3 E^{i k} E_k^j\,,
\eqno(30)
$$
where $I_1(\ie) = E_k^{\,\,\,k}$ and $I_1(\ie^2) = E_{k l} E^{k l}$.
When $\nu_{1, 2, 3}$ are zero, we obtain (2), while $\si_{i j}$
coincides with (4) in the weak field limit.

The problem of corrections to the linear dislocation solutions
has been actively investigated by various methods. For instance, the
Refs.[20, 21, 35, 36], as well as a review of them in [31], are to be
mentioned here. The Refs. [31, 32, 37] are also useful as the sources of
information about others approaches to the non-linear dislocation problems.
Moreover, in [38] second order corrections have been obtained for a wedge
disclination solution. All the calculations in the refs. mentioned have been
done assuming the constitutive relation in the Murnaghan's form to account
for all the quadratic contributions properly. The Ref.[7] is concerned with
the non-conventional approach to the screw dislocation [1, 2]. However the
constants $\nu_{1, 2, 3}$ are taken zero in [7].

\subsection{Vanishing torsion ?}

Before to conclude the paper, the Eq.(14) has been proposed in the Sec.3
to avoid the contradiction between $p$ (10) and $\si_{3 3}=\nu(\si_{1 1}
+\si_{2 2})$ for the edge dislocation. In their turn, Sec.4.2, 4.3,
are  to extend (14) to the Eq.(28), which is concerned  with the
Riemann--Christoffel picture. Therefore, it is appropriate to check, whether
the solutions considered above are consistent with the requirement of
zero torsion. Obviously, the torsion can not be asked to vanish everywhere,
because thus defects will drop out. It seems sufficient, since our
description should approximate a more adequate Riemann--Cartan situation,
to ask about the torsion which is zero, at least, at $\kappa\rho \ll
1$, $\,0 < R_{core} < \rho < R_{exterior} < \infty $.

First, it is appropriate to remind the situation [12] with the screw
defect, which is more transparent. It is known how to calculate the torsion
tensor [24] (see also (21)), and so we obtain
from (4) and (7) for its single non-zero component:
$$
{{\cal T}^3}_{,\,1 2}\,=\,\frac{\,1\,}2\,\left[2(\cd_1\om_2\,+\,\cd_2\om_1)
-\Dl f \right]\,,
\eqno(31)
$$
where the modified stress function $f$ fulfills (18). There are two auxiliary
functions $\om_1$ and $\om_2$ in (31) to account for the antisymmetric
part of distortion:
$$ \bphi\,=\,\frac{\,1\,}2\,\left( \begin{array}
{ccc} 0  &    0 & \cd_2 f-2\om_1\\
      0  &    0 & -\cd_1 f+2\om_2\\
 \cd_2 f+2\om_1 & -\cd_1 f-2\om_2 & 0
\end{array}\right)\,.  $$
In the
(anti-)plane problem we choose $\om_{1}$, $\om_{2}$ so that the
distortion components $\phi_{3 1}$ and $\phi_{3 2}$ are zero (for the defect
along $Ox_3$) [8]. Then (31) is reduced to
$$
{{\cal T}^3}_{,\,1 2}\,=\,-\,\Dl f\,.
\eqno(32)
$$
In the axisymmetric case, the solution to (18) can be written as
$f = (b/2\pi) K_0(\kappa\rho)$ [1, 2]. At $\kappa\rho \ll 1$,
this solution
results approximately in $(b/2\pi) \log (1/\rho) + const$, i.e. in the
Prandtl's stress function, while the corresponding value of
${{\cal T}^3}_{,\,1 2}$ (32) is zero, since we get
$b\,\dl(x_1)\dl(x_2)$ at $\rho\ne 0$. At $\kappa\rho \gg 1$, the
torsion ${{\cal T}^3}_{,\,1 2}$ approximately vanishes by (18), (32),
since the Bessel function decays exponentially. Provided the entire
$f$ is used to consider the torsion component at $\kappa\rho \ll 1$,
one gets:
$$
{{\cal T}^3}_{,\,1 2}\,\simeq\,-\,\frac{\,b\,}{2 \pi}
\kappa^2 \left( const\,-\,\log(\kappa\rho)\right)\,.
$$
The latter still vanishes because $\rho$ is limited
$R_{core} < \rho < R_{exterior}$ and $\kappa\to 0$. However,
$\,{{\cal T}^3}_{,\,1 2}$ (32) is not zero in the intermediate region
$\kappa\rho\simeq 1$, thus violating our Riemannian interpretation.

Is it possible to make ${{\cal T}^3}_{,\,1 2}$ zero by an appropriate choice
of $\om_1$, $\om_2$? Formally, it can be done relaxing the requirement
$\phi_{3 1}=\phi_{3 2}=0$. But therefore the framework of the
(anti-)plane problem is left in favour of a 3-dimensional
consideration. Notice, that the torsion components have not been
considered in [1, 2] for the screw dislocation as we just did. This
has influenced, for instance, the Burgers vector component $b_3$ found
in [1] as $b/2$ instead of $b$ owing to the neglection of
$\phi_{3 1}=\phi_{3 2}=0$ at $\kappa\to 0$.

Now let us turn to the edge dislocation which is more
complicated. Here the torsion components are:
$$
{{\cal T}^1}_{,\,1 2}\,=\,\frac{\,1\,}2\,\left[2\,\cd_1\om\,+\,(1-\nu)
\cd_2\Dl f\,+\,\nu\,\kappa^2\cd_2 f\right]\,,
$$
$$
\eqno(33)
$$
$$
{{\cal T}^2}_{,\,1 2}\,=\,\frac{\,1\,}2\,\left[2\cd_2\om\,-\,
(1-\nu)\cd_1\Dl f\,-\,\nu\,\kappa^2\cd_1 f\right]\,,
$$
where $\om$ is to account for the antisymmetric part in the distortion
component $\phi_{1 2}$ (the normalization of $f$ corresponds to (7)).
The dependence on $\kappa$ is present in (33), because
$p$ (16) has been used. When $f$, ${{\cal T}^1}_{,\,1 2}$, and
${{\cal T}^2}_{,\,1 2}$ are fixed, the function $\om$ can be found
from (33) provided, say, the integrability condition
$(\cd_1\cd_2-\cd_2\cd_1)\om=0$ is implemented, i.e. the equation
$$
(1-\nu)\Dl\,\Dl f\,+\,\nu\,\kappa^2\,\Dl f\,=\,
2\left(\cd_2 {{\cal T}^1}_{,\,1 2}\,-\,\cd_1 {{\cal T}^2}_{,\,1 2} \right)\,,
\eqno (34)
$$
holds.

In the formal limit $\kappa\to 0$ the choice
$$
\om\,=\,\frac{\,b\,}{2\pi}\frac{\,x_1\,}{\rho^2}
$$
ensures that (33) can be fulfilled with
$${{\cal T}^1}_{,\,1 2}\,=\,b\,\dl(x_1)\dl(x_2)\,,\qquad
{{\cal T}^2}_{,\,1 2}\,=\,0\,,
\eqno(35)
$$
and $f=(b/4\pi)(1-\nu)^{-1} \cd_2(\rho^2\log\rho)$. The torsion,
Eq.(35), is given, the integrability Eq.(34) acquires the form at
$\kappa\to 0$:
$$
(1-\nu)\,\Dl\Dl f\,=\,2 b\,\dl(x_1)\cd_2\dl(x_2)\,,
$$
thus prescribing the limiting form of the modified stress function.

Taking into account the explicit solution $f$ (20), and using (33),
(34), we can also calculate the torsion components ${{\cal T}^1}_{,\,1
2}$, ${{\cal T}^2}_{,\,1 2}$ at arbitrary $\kappa$. Indeed, let us take
${{\cal T}^2}_{,\,1 2}=0$ to express $\om$. Then we obtain
${{\cal T}^1}_{,\,1 2}$:
$$
{{\cal T}^1}_{,\,1 2}\,=\,
\frac{\,b s\,}{4 \pi}\,\left((1 - \nu)\,\Dl\Dl {\cal F}\,+\,
\nu \kappa^2 \Dl {\cal F} \right)\,=$$
$$=\,-\frac{\,b\,}{4 \pi}\,\mu\,\nu\,\left( \Dl
{\cal F}\,-\,\frac{\,\kappa^2\,}a {\cal F} \right)\,,
$$
where the Eq.(19) has been used to exclude $\Dl\Dl$. Using ${\cal F}
(\rho)$ (20) explicitly, we obtain further:
$$
{{\cal T}^1}_{,\,1 2}\,=\,
\frac{\,b\,}{4 \pi}\,\frac{\,\mu \kappa^2\,}{1 - \nu}\,
\left((1 + \nu)\,{\cal F}(\rho)\,+\,
2 \nu\,K_0({\cal M} \rho) \right)\,.
\eqno(36)
$$
At ${\cal M}\rho, {\cal N}\rho \ll 1$, the R.H.S. of (36) is also
vanishing due to the same reasons, as it happens for
${{\cal T}^3}_{,\,1 2}$ (32) in the analogous situation. At
${\cal M}\rho, {\cal N}\rho \gg 1$, $\,{{\cal T}^1}_{,\,1 2}$
decays as linear combination of $\rho^{-1/2}\sin({\cal N}\rho)$
and $\rho^{-1/2} \cos({\cal N}\rho)$, i.e. not so fast as ${{\cal
T}^3}_{,\,1 2}$ (32) does.

Is it possible to make ${{\cal T}^1}_{,\,1 2}$, ${{\cal T}^2}_{,\,1 2}$
zero by adjusting $\om$? The integrability condition for $\om$
can be written, without the use of (16), as follows:
$$ (1-a)\Dl\,\Dl f\,+\,a\,\Dl p\,=\,0 $$
Then, formally, at finite $s$ the parameter $p$, i.e. the stress
component $\sigma_{3 3}$,
is zero by (7), (17). Therefore the strain component $E_{3 3}$ is
$$
E_{3 3}\,=\,-\frac{a}{2\mu}(\sigma_{1 1}\,+\,\sigma_{2 2})\,=\,
\frac{a}2\,\Dl f\,.
$$
If $E_{3 3}$ is still zero (as at $\kappa\to 0$), the function
$f$ is also zero by (16). Otherwise the two-dimensional consideration
is no longer sufficient.

Let us sum up this section. Here we have obtained the torsion components
expressed through the stress functions. The results of the Sec.3
have been used to establish that there exists a single limiting
situation, i.e. the limit $R_{core} < \rho < R_{exterior}$,
$\kappa\to 0$,
when the modified stress functions can be replaced approximately by the
conventional ones, and the torsion becomes zero. This would imply that the
Einsteinian interpretation of the Eq.(14) is valid only in that limit,
while the modified stress functions themselves for others $\kappa$,
$\rho$ should be discarded at the present stage.

\section{DISCUSSION}

Thus, in the present paper we have started with the demonstration of
the fact that the conventional linear solution for the edge dislocation
does not fit in the gauge model [1, 2] (specifically, its
translational sector). The Kr\"oner ansatz for stresses is used in
[1, 2] to fulfil identically the equilibrium equations of the medium.
It is essential that the corresponding parametrizing functions, i.e. the
(modified) stress functions, are to be found from the translational
gauge equation. In its turn, the form of the translational gauge equation
[1, 2] is dictated by the Lagrangian quadratic in the $T(3)$-gauge field
strength. As a special example, in [1, 2] the Kr\"oner ansatz has been
chosen in that form [19], which leads in conventional theory to the
Airy's stress function of the edge dislocation along third axis.
The present paper displays that in this case the gauge equation
[1, 2] constraints the modified stress functions so that the resulting
stress component $\si_{3 3}$ arises as $\nu^{-1}(\si_{1 1}+ \si_{2
2})$ instead of $\nu (\si_{1 1}+\si_{2 2})$ at $\kappa\to 0$,
i.e.  $\si_{3 3}$ is not of the edge dislocation.

The ansatz itself looks more valuably for the problem in question, and
therefore the contradiction found can be eliminated by another choice of
the master translational gauge equation. Namely, the use of the double curl
differential operator in its L.H.S. leads to the replacement
$a\longleftrightarrow 1-a$, or, equivalently, $\nu\longleftrightarrow
\nu^{-1}$. More precisely, both the linear gauge equations in question
(our double curl Eq.(14)and that from [1, 2]), provided they are
written through the general tensor of stress functions, pass each into
other just simply by inverting the Poisson ratio $\nu$. Eventually,
both the standard linear solutions for dislocations are possible with
the new gauge equation, though in the formal limit $\kappa\rho \ll
1$ ($\kappa\to 0$, $\rho$ is finite).

From a more formal point of view on Lagrangian field-theoretical modelling
of conventional (and non-conventional) defects, the newly proposed Eq.(14)
looks suitable for the replication of the dislocation solutions.
Therefore, it is worth to be realized that the double curl in the L.H.S.
of (14) can be considered not only formally, as a nice trick, but also as
linearization of a three-dimensional Einstein tensor. Moreover,
a non-linear constitutive relation `stress--strain' can be thought of
instead of the R.H.S. of (14). Thus, (14) can be treated as a weak
field limit of a non-linear Einstein-type Eq.(29), which is the most
straightforward extension of (14). The Eq.(29) uses the framework of
torsion--free Riemannian geometry, and it can be derived by Lagrangian
method.

The torsion components can also be expressed through the stress functions
by means of the relation `strain--stress' and the Kr\"oner ansatz. The
direct calculation in the linear approximation shows us that these
components are not zero, though vanishing can be asked
for a special choice of the stress functions. In our situation,
Sec.3, vanishing occurs in the particular limit
$\kappa\rho\ll 1$, when the
modified stress functions coincide approximately with the Airy's and
Prandtl's ones. This simply means that the torsion-less interpretation
behind (28) is valid for (14) just at $\kappa\rho\ll 1$. (Notice,
that non-linear constitutive relation has to be used to re-express
torsion in general situation).

Strictly speaking, the results concerning the torsion rather imply a
necessity to extend the framework and to pass to
the Riemann--Cartan formalism, which will allow for a non-trivial torsion.
However, the Eq.(28) itself still looks appropriate
in the following sense. As soon as the matter is concerned with the
defects sufficiently separated each from other, the
torsion can be required as localized
inside tubes confining the defects. Outside such tubes (i.e. for $\rho >
R_{core} > 0$) but not very faraway from the cores ($\rho < R_{exterior}
< \infty$), the Eq.(28) can approximately be valid for the purely
translational defects at $\kappa\rho\ll 1$.
Indeed, the Eq.(14) with the solutions found capture
properly the required conventional stress functions and
can be considered as a first step of an iterative scheme at
$\kappa\to 0$, provided (28) would dictate next corrections, valid
outside $R_{core}$.

Recall that the stress function approach [20, 21] postulates
vanishing of the Einstein tensor, while the torsion (i.e. the
density of dislocations) is presumed $\dl$-like to obtain the first order
approximation solutions. This picture corresponds to teleparallel geometry.
On the contrary, in our case the source in the Einstein equation is
present, but it is smeared. Hence, the first order
solutions appear only in the special limit $\kappa\to 0$.

It is the Hilbert--Einstein Lagrangian which provides the Einstein
tensor as the L.H.S. of the gauge equation. However it requires a
modification of the geometric picture [1, 2]. Practically, the use of the
``coordinate'' representation seems to be appropriate. Besides, now it is
more clear, how the matter Lagrangian ${\cal L}_\xi$ affects the
linearized model, e.g., it leads to the limited modified stress
functions instead of the increasing conventional ones. Although we
have started with a rather special problem of modification of the
$\si_{3 3}$-component, it seems, eventually, that the ideas
discussed here should be valid for the whole $\ISO$-model also.

As the immediate implication of the present consideration, let us mention
the following one. The second order corrections to the conventional screw
dislocation, which are implemented by [1, 2], have been obtained in [7].
As far as $\L$ (3) is quadratic in derivatives of the gauge potential,
the L.H.S. of the gauge equation is of first order in second derivatives.
Therefore, it remains the same for
both perturbative steps in [7]. But the Lagrangian $\L$ (3) is shown to be
fairly inconsistent when replicating the edge dislocation, while the
Hilbert--Einstein one removes the problem. Therefore
usage of $\L$ suggested here could influence the results of
[7], because the L.H.S. of the corresponding Einstein-type equation is
non-linear. Moreover, the use in [7] of the constitutive relation without
the Lam$\acute e$ constants of third order could be restrictive.

Generally speaking, both the Lagrangians, the Hilbert--Einstein
and the quadratic, can occur in the gauge--translational models.
The matter should be to specify a proper range of problems for each of
them. The consideration presented is to show that the
Hilbert--Einstein Lagrangian is rather appropriate when replicating the
conventional dislocations, but further investigations are needed to go
beyond the weak field. In its turn, the Lagrangian (3) can
be more appropriate in such problems as [3--5]. Besides, the
Refs.[40] should be mentioned where a Lagrangian quadratic in the
translational field strength has been used in more general form to
study a non-gravitational structure on space-time.

To summarize, a non-linear Lagrangian $T(3)$-gauge
model is proposed which admits at $\kappa\rho\ll 1$,
the conventional dislocation solutions of the
linear isotropic (incompatible) elasticity. The calculation presented
can be considered as a first order approximation. Incorporation of
second order contributions to the gauge equation, and/or extension to
the $\ISO$-case to include the torsion are needed to extend the
present consideration, and to decide about further perspectives of the
$\ISO$-gauging for defects modelling. With regard to the great
attention to the $T(3)$-gauge models with the quadratic $\L$, it is
hopeful that the work presented would serve to a more adequate understanding
of the problem discussed.

A historical remark: more superficial versions of the critical
content of the Sec.2 and of the correction discussed in Sec.3 can be
found, respectively, in [41] and [42].

\medskip
\subsection*{ACKNOWLEDGEMENTS}
\medskip

The author is grateful for warm hospitality to the Center for Theoretical
Physics in Warsaw, where the paper has been completed. It is a real
pleasure to thank Prof.L.A.Turski for his kind invitation to visit the
Center and for his numerous and valuable discussions on defects in
condensed matter. The author thanks Prof.V.M.Babich for discussing the
Sec.3, and Profs.F.W.Hehl and A.E.Romanov for reading the manuscript. The
research described has been partially supported by the Russian Foundation
for Fundamental Research Projects No. 96--01--00807, 98--01--00313 and by
the Grant from the J. Mianowski Foundation for Science Promotion (Poland).

\end{document}